# The first application of high-order Virial equation of state and *ab initio* multi-body potentials in modeling supercritical oxidation in jet-stirred reactors

Mingrui Wang[a], Ruoyue Tang[a], Xinrui Ren[a], Hongqing Wu[a], Ting Zhang[a], Song Cheng[a,b*]

[a] *Department of Mechanical Engineering, The Hong Kong Polytechnic University, Kowloon, Hong Kong SAR, China*
[b] *Research Institute for Smart Energy, The Hong Kong Polytechnic University, Kowloon, Hong Kong SAR, China*

## Abstract

Supercritical oxidation processes in jet-stirred reactors (JSR) have been modeled based on ideal gas assumption. This can lead to significant errors in or complete misinterpretation of modeling results. Therefore, this study newly developed a framework to model supercritical oxidation in JSRs by incorporating *ab initio* multi-body molecular potentials and high-order mixture Virial equation of state (EoS) into real-fluid conservation laws, with the related numerical strategies highlighted. With comparisons with the simulation results based on ideal EoS and the experimental data from high-pressure JSR experiments, the framework is proved to be a step forward compared to the existing JSR modeling frameworks. To reveal the real-fluid effects on the oxidation characteristics in jet-stirred reactors, simulations are further conducted at a wide range of conditions (i.e., temperatures from 500 to 1100 K and pressures from 100 to 1000 bar), the real-fluid effect is found to significantly promote fuel oxidation reactivity, especially at low temperatures, high pressures, and for mixtures with heavy fuels. The significant influences of real-fluid behaviors on JSR oxidation characteristics emphasize the need to adequately incorporate these effects for future modeling studies in JSR at high pressures, which has now been enabled through the framework proposed in this study.

*Corresponding author.
Email: songcheng@polyu.edu.hk
Song Cheng
Phone: +852 2766 6668

## Novelty and Significance Statement

This study is the first time that high-order Virial EoS and *ab initio* multi-body potentials are applied to supercritical oxidation modelling in jet-stirred reactors. This is enabled through a newly developed framework coupling *ab initio* intermolecular potentials, high-order mixture Virial EoS, and real-fluid conservation laws. Through a combination of validations and simulations (with results and methods detailed below), we further revealed the strong impact of real fluid effects on supercritical oxidation processes from fundamental perspectives, and compared these effects over wide ranges of temperature, pressure and fuels. Finally, recommendations have also been made for future modeling and experimental studies in JSR at high pressures.

## Author Contributions

M.W. and S.C. conceived and designed the framework and performed the calculations, with technical inputs from R.T., X.R., H.W., and T.Z. All authors contributed to writing the paper.

# 1. Introduction

Supercritical combustion holds practical potential in transportation and power generation industries owing to its advantages in enhanced power output, improved thermal efficiency and high fuel economy [1, 2]. Particularly, low-temperature supercritical combustion (T<1200 K) is favored for lower heat loss, lower pollution emission, and wider applications than high-temperature combustion [3]. This trend becomes particularly obvious for power engines applied in electricity sectors, aviation and aerospace vehicles where power density is highly sought-after that batteries become not applicable. As a result, the power density of combustion and propulsion systems has increased exponentially over the last thirty years. Modern on-road heavy-duty engines, gas turbines, and rocket engines can easily operate at pressure levels ranging from 100 to 300 atm [4], where the working fluid becomes supercritical.

In supercritical fluids, intermolecular interactions (especially multi-body interactions) become important and are significantly augmented, leading to considerable drifts in both molecular collision frequency and collisional cross section [5]. These changes deviate molecule distributions from the Maxwell-Boltzmann distribution. As such, supercritical fluid behaves differently from ideal fluids, sharing, for instance, the characteristics of liquid (e.g., high density) and gas (e.g., high diffusivity) [6] with high compressibility factors [7]. These shifted fluid behaviors will affect combustion properties profoundly. For combustion processes at low temperatures, molecular behaviors are expected to be more complicated, as various radicals and intermediates, in addition to the stable species, are abundantly generated [8] with longer residence time than those produced at high temperatures. It's necessary to gain an accurate and quantitative understanding of such molecular behaviors and their respective influences during supercritical combustion, which is, however, absent from previous studies.

Supercritical combustion simulation presents challenges to numerical modelling which should adequately incorporate real-fluid thermodynamics, real-fluid chemical kinetics, and real-fluid transport properties. In homogeneous systems, e.g., jet-stirred reactors (JSR) [9, 10] and rapid compression machines [11, 12], optimizing fuel chemistry model (usually by changing reaction kinetic parameters) against indirect experimental measurements (e.g., species profiles and ignition delay times) has been adopted as one means to counteract the real-fluid effects. However, in these attempts, the ideal gas assumption is still maintained, making simulations only applicable to the experiments used for model validation, which lacks physical significance. Another way is to correct the real-fluid thermochemical properties by adopting a real-fluid equation of state (EoS), primarily the empirical cubic EoS, such as Redlich- Kwong (RK) EoS [13, 14], Peng-Robinson EoS [15-17], and Soave-Redlich-Kwong EoS [18]. In spite of simplicity and convenience, cubic EoS is empirical and cannot reveal the physical nature of the real-fluid effect from the fundamental viewpoints of intermolecular interactions.

On the other hand, the Virial EoS [19] was developed based on the real-fluid partition function theory in statistical mechanics, where fluid non-ideality is directly quantified using Virial coefficients that are directly calculated from intermolecular potentials. These Virial coefficients physically represent the intermolecular interactions in the fluid, which are computed from pre-determined intermolecular potentials. A recent study from the authors' group has shown that high-order Virial EoS developed from *ab initio* intermolecular potentials exhibits remarkable superiority to cubic EoS, impressively replicating the thermodynamic properties of various systems at supercritical conditions [20]. In that study, a new framework was also proposed to model the autoignition process at trans- and super-critical conditions where significant changes in autoignition reactivity were observed. Despite the successful demonstration in [20], Virial EoS has not been applied to JSR modeling, let alone those based on high-order Virial EoS and *ab initio* intermolecular potentials.

With such awareness, this study aims to: (1) develop a robust framework for trans- and super-critical JSR modeling that couples *ab initio* molecular potentials, high-order mixture Virial EoS, and real-fluid oxidation conservation laws; and, based on which, (2) investigate the real-fluid effects on oxidation processes, and how these effects will change cross various temperatures, pressures and fuels.

# 2. Methodologies

## 2.1 High-order mixture Virial EoS

The $N^{th}$-order Virial EoS can be written as:

$$Z = \frac{P\bar{v}}{RT} = \frac{\bar{v}}{\bar{v}^{IG}} = 1 + \frac{B_2}{\bar{v}} + \cdots + \frac{B_N}{\bar{v}^{N-1}} \tag{1}$$

where $T$, $P$, $Z$, and $R$ denote the temperature, pressure, compressibility factor and universal gas constant, respectively. $\bar{v}$ and $\bar{v}^{IG}$ are the molar volume of the real fluid and ideal gas respectively. $B_2$, $B_3$, …, $B_N$ represent the second, third, …, and $N^{th}$-order Virial coefficients, corresponding to intermolecular interactions between two molecules, three molecules, …, and $N$ molecules, respectively.

When Eq.1 is used for pure substances or mixtures, the related Virial coefficients are accordingly named pure-substance Virial coefficients or mixture Virial coefficients. Mixture Virial coefficients can be calculated by the mixing rule [19]. For instance, the $N^{th}$-order mixture Virial coefficient for an R-S binary mixture is computed by:

$$B_N(T, X_R, X_S) = \sum_{r+s=N;\ r,s\geq 0} \frac{N!}{r!\,s!} X_R^r X_S^s B_{rs} \tag{2}$$

where $B_{rs}$ denotes the Virial coefficient for $r$ molecules of substance $R$ and $s$ molecules of substance $S$, while $X_R$ and $X_S$ are mole fractions of each component. If $r$ or $s$ equals 0, $B_{rs}$ represents the $N^{th}$-order pure-substance Virial coefficients of the substance $R$ or $S$ respectively. If $r$ and $s$ are non-zero, $B_{rs}$ is named the $N^{th}$-order cross Virial coefficients, related to the intermolecular interaction between substances $R$ and $s$. As the pure-substance Virial coefficient of a specific substance is determined by only temperature, the mixture Virial coefficient is a function of temperature and mole fractions in essence.

The method of generating high-order mixture Virial EoS based on *ab initio* intermolecular potentials for an arbitrary system including radicals has been recently proposed and validated in [20]. Therefore, this method will be briefly introduced here, with the detailed theories summarized in Chapter 2 of the Supplementary Material. The whole logic is illustrated in Fig. 1, mainly consist of three steps:

(1) Determining pure-substance Virial coefficients. The high-order Virial coefficients derived from *ab initio* intermolecular potentials are adopted, as summarized in Table S1, which cover the major stable species in typical combustion systems (e.g., $CO_2$, $H_2O$, $N_2$, $O_2$) and the important fuel derivatives (e.g., $CH_4$, $CH_3CHO$, $H_2$). Except for those having been calculated before, the low-order Virial coefficients of pure substances including radicals and intermediates are predicted by the principle of corresponding state [21, 22], which expresses the Virial coefficient as a function of critical temperature and pressures. If necessary, critical points of an arbitrary substance including radicals can be estimated by the Joback group contribution (JGC) method [23].

(2) Predicting cross Virial coefficients. The low-order cross Virial coefficients can be similarly estimated by the principle of corresponding state [21, 22, 24], while the high-order ones will be directly predicted by the "new combining methods" [20] from *ab initio* high-order pure-substance Virial coefficients. The new combining methods have already been validated in [20], demonstrating excellent accuracy with those calculated *ab initio*.

(3) Based on the mixing rule (Eq. 2), mixture Virial coefficients can be calculated from pure-substance Virial coefficients and cross Virial coefficients, thereby giving high-order mixture Virial EoS.

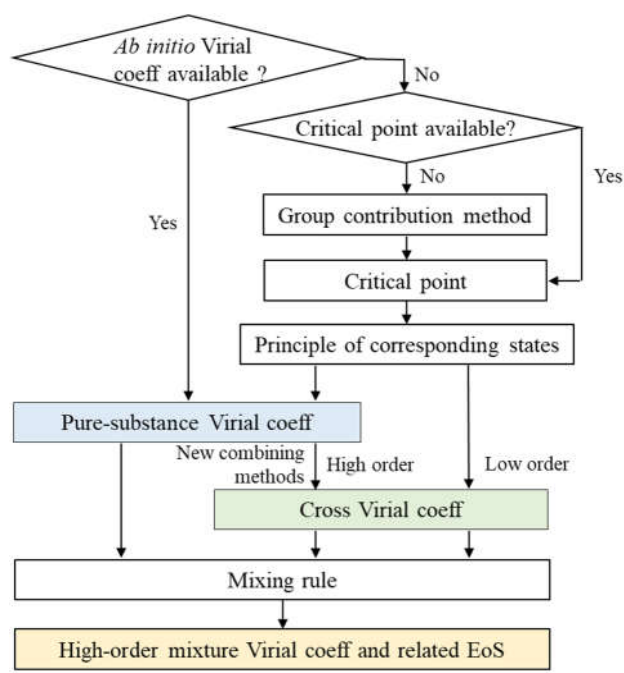

**Figure 1. The logic for deriving high-order mixture Virial EoS.**

The high-order mixture Virial EoS has been validated [20] by experiments and proved to be accurate, outperforming the ideal EoS and traditional RK EoS in predicting thermochemical properties for a wide range of species. For a detailed validation and comparison process, please refer to Chapter 3 of the Supplementary Material.

## 2.2 JSR model based on high-order Virial EoS

### 2.2.1 Governing equations

Based on the high-order mixture Virial EoS developed, we then propose a new framework to simulate supercritical oxidation in jet-stirred reactors. Several assumptions are made for an *I*-component system:

(1) The steady-state flow has the same steady rate everywhere and has uniform properties in the direction perpendicular to the flow.
(2) The steady-state temperature and pressure are known and given, and thus there is no need to consider energy balance issues.

The mass conservation law of the $i$ species in the reactor is written as:

$$\dot{m} = VM_i w_i + \dot{m}_{in} Y_{i,in} - \dot{m}_{out} Y_{i,out} \qquad i = 1,2\ldots,I\ species \qquad (3)$$

where $M_i$, $w_i$ and $Y_{i,in}$ are the molecular weight, net production rate and mass fraction of the $i$ species, respectively. $V$ is the volume of the reactor. $\dot{m}$ is the mass change rate in the reactor, and $\dot{m}_{in}$ and $\dot{m}_{out}$ are the mass flow rates at the inlet and outlet, respectively, of the reactor. For steady-state flow: $\dot{m} = 0, \dot{m}_{in} = \dot{m}_{out}$. Thus, Eq.3 is rewritten as:

$$VM_i w_i + \dot{m}(Y_{i,in} - Y_i) = 0 \qquad (4)$$

In addition, the mole fraction $X_i$ and mass fraction $Y_i$ have the following relationship:

$$Y_i = \frac{X_i M_i}{\bar{M}} \qquad (5)$$

where $\bar{M}$ denotes the average molecular weight.

Then, combining Eq.4 and Eq.5, the mass conservation for each species becomes:

$$\frac{V w_i}{\dot{m}} + \frac{X_{i,in}}{\bar{M}_{in}} - \frac{X_i}{\bar{M}} = 0 \qquad i = 1,2\ldots,I\ species \qquad (6)$$

According to the Gibbs Phase Rule [25], the steady state of this system is determined by temperature, molar volume, and the mole fraction of each substance. As temperature is constant and has been given, there are $I + 1$ unknown variables ($\bar{v}, X_1 \ldots X_I$) need to be solved, while there are only $I$ mass conservation equations given (Eq.6). Therefore, the $N^{th}$-order mixture Virial EoS is used as another governing equation (Eq.7). In addition, to keep the sum of mole fractions at 1, one of mass conservation equations in Eq.6 are replaced by Eq.8. With these treatments, the final set of governing equations becomes:

$$f_{eos}(\bar{v}, X_1 \ldots X_I) = -\frac{P\bar{v}}{RT} + 1 + \sum_k \frac{B_k}{\bar{v}^{(k-1)}} = 0 \qquad k = 2,3,\ldots N \qquad (7)$$

$$f_{sum}(\bar{v}, X_1 \ldots X_I) = \sum_i X_i - 1 = 0 \qquad (8)$$

$$f_{mas,i}(\bar{v}, X_1 \ldots X_I) = \frac{V w_i}{\dot{m}} + \frac{X_{i,in}}{\bar{M}_{in}} - \frac{X_i}{\bar{M}} = 0 \qquad i = 1,2\ldots,I-1\ species \qquad (9)$$

Eq.7-9 represent $I + 1$ algebraic equations, a classical two-point boundary value problem.

2.2.2 Numerical methods

According to previous ideal gas combustion modelling in JSR [26], the Newton-Raphson method and the related optimized methods [27] are proven to be reliable and will be adopted to find steady-state solutions in this study. For better and faster convergency, a $(I + 1) \times (I + 1)$ Jacobian matrix is indispensable:

$$\text{Jacobian matrix} = \begin{bmatrix} \frac{\partial f_{eos}}{\partial \bar{v}} & \frac{\partial f_{eos}}{\partial X_1} & \cdots & \frac{\partial f_{eos}}{\partial X_l} & \cdots & \frac{\partial f_{eos}}{\partial X_I} \\ \frac{\partial f_{mas,1}}{\partial \bar{v}} & \frac{\partial f_{mas,1}}{\partial X_1} & \cdots & \frac{\partial f_{mas,1}}{\partial X_l} & \cdots & \frac{\partial f_{mas,1}}{\partial X_I} \\ \cdots & \cdots & \cdots & \cdots & \cdots & \cdots \\ \frac{\partial f_{mas,i}}{\partial \bar{v}} & \frac{\partial f_{mas,i}}{\partial X_1} & \cdots & \frac{\partial f_{mas,i}}{\partial X_l} & \cdots & \frac{\partial f_{mas,i}}{\partial X_I} \\ \cdots & \cdots & \cdots & \cdots & \cdots & \cdots \\ \frac{\partial f_{mas,I-1}}{\partial \bar{v}} & \frac{\partial f_{mas,I-1}}{\partial X_1} & \cdots & \frac{\partial f_{mas,I-1}}{\partial X_l} & \cdots & \frac{\partial f_{mas,I-1}}{\partial X_I} \\ 0 & 1 & \cdots & 1 & \cdots & 1 \end{bmatrix}$$

The partial differentials in the Jacobian matrix are newly derived, expressed as:

$$\frac{\partial f_{eos}}{\partial \bar{v}} = -\frac{P}{RT} - \sum_k \frac{(k-1)B_k}{\bar{v}^k} \tag{10}$$

$$\frac{\partial f_{eos}}{\partial X_l} = \sum_k \left( \frac{1}{\bar{v}^{(k-1)}} \frac{\partial B_k}{\partial X_l} \right) \qquad l = 1,2\ldots,I \text{ species} \tag{11}$$

$$\frac{\partial f_{mas}}{\partial \bar{v}} = \frac{V}{\dot{m}} \frac{\partial w_i}{\partial \bar{v}} \tag{12}$$

$$\frac{\partial f_{mas,i}}{\partial X_l} = \begin{cases} \frac{V}{\dot{m}} \frac{\partial w_i}{\partial X_i} - \frac{\overline{M} - X_i M_i}{\overline{M}^2} & l = i \\ \frac{V}{\dot{m}} \frac{\partial w_i}{\partial X_l} + \frac{X_i M_l}{\overline{M}^2} & l \neq i \end{cases} \tag{13}$$

Note that in the Jacobian matrix, the net production rate $w_i$ and its differential with respect to the mole fraction, $\partial w_i / \partial X_i$, are directly calculated by the CANTERA software [28], whereas its differential related to molar volume, $\partial w_i / \partial \bar{v}$, can not be obtained directly. Eq.14 establishes a relationship between $\partial w_i / \partial \bar{v}$ and $\partial w_i / \partial \rho_n$, with the latter one computed by CANTERA directly ($\rho_n$ represents the molar density).

$$\frac{\partial w_i}{\partial \bar{v}} = -\frac{1}{\bar{v}^2} \frac{\partial w_i}{\partial (n/V)} = -\frac{1}{\bar{v}^2} \frac{\partial w_i}{\partial \rho_n} \tag{14}$$

where $n$ is the total amount of all substances in the reactor.

Previous studies found that inlet conditions ($\bar{v}^0, X_1^0 \ldots X_I^0$) might not be sufficient as the Newton-Raphson solver and the governing equations are sensitive to initial values [26, 29]. Thus, the inlet condition will be first input into CANTERA, and subsequently ideal equilibrium properties ($\bar{v}^E, X_1^E \ldots X_I^E$) will be computed, serving as the initial values for the Newton-Raphson solver [29].

For better illustration, the supercritical JSR modelling framework is depicted in Fig. 2, with numerical strategies highlighted. The framework consists of a "Virial Module", a "CANTERA Module", and a "Solver Module", distinguished by blue, green, and yellow translucent backgrounds, respectively. To begin with, an appropriate chemical mechanism needs to be selected, and the "Virial Module" can accordingly generate a set of high-order Virial EoS, with the partial differential of Virial coefficients with respect to mole fractions calculated. Next, the inlet conditions for simulation will be input into the "CANTERA Module", with ideal equilibrium properties, net production rates, and the partial differential of net production rates calculated. At the final stage, ideal equilibrium properties will serve as the initial condition, while the other variables output from the "Virial Module" and "CANTERA Module" will be substituted into combustion governing equations and generate a Jacobian matrix. Numerical solutions will be obtained from the "Solver Module" based on the Newton-Raphson method.

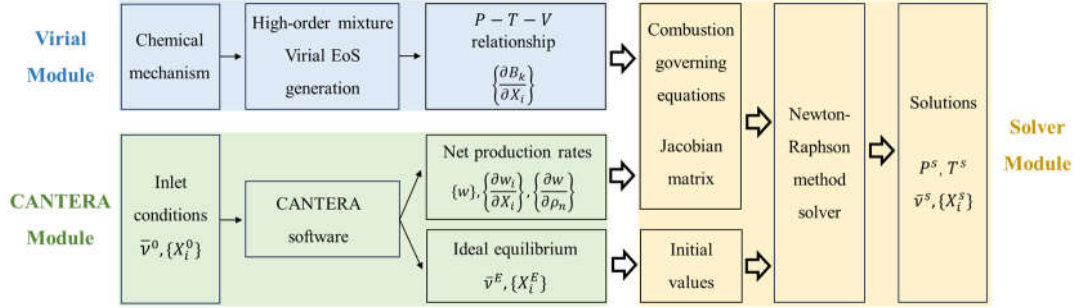

**Figure 2.** Supercritical oxidation modeling framework based on high-order mixture Virial EoS and *ab initio* intermolecular potential for jet-stirred reactors.

### 2.2.3 Chemical kinetic mechanism

As discussed with Fig. 2, a chemical kinetic mechanism is needed as the starting point. The NUIGMech1.1 mechanism [30] is adopted in this study, which has been validated comprehensively for the fuels to be investigated in this study.

### 2.2.4 Inlet conditions

To validate the framework developed, supercritical experimental data obtained in a JSR for three different mixtures are adopted for comparison, and the inlet conditions are illustrated in Table 1.

**Table 1** Inlet conditions selected for supercritical JSR modeling with experimental data available for comparison.

| Case | P (atm) | T (K) | Mixture | Dilution | Equivalence ratio | Experiment ref |
|---|---|---|---|---|---|---|
| 1 | 100 | 534-889 | $CH_3OH/O_2/N_2$ | 95.17-96.90 | 0.1-9.0 | [31] |
| 2 | 100 | 497-888 | $CH_3OCH_3/O_2/N_2$ | 88.27-96.43 | 0.175-1.72 | [32] |
| 3 | 100 | 569-895 | $C_3H_8/O_2/N_2$ $C_3H_8/O_2/N_2/CO_2$ | 91.63-98.11 | 0.2-1.75 | [33] |

To explore the real-fluid effects over wider conditions, and to establish consistent comparisons between different fuels, fuel loadings conditions and thermodynamic conditions, simulation is conducted for the three mixtures in Table 1 at a wider temperature and pressure range, with fixed diluent ratio and equivalence ratio. The inlet conditions of JSR for these investigations are illustrated in Table 2.

**Table 2** Inlet conditions for supercritical JSR modeling at consistent and wider conditions.

| Case | P (bar) | T (K) | Mixture | Dilution | Equivalence ratio |
|---|---|---|---|---|---|
| 4 | | | $CH_3OH/O_2/N_2$ | | |
| 5 | 100-1000 | 500-1100 | $CH_3OCH_3/O_2/N_2$ | 90 | 0.1 |
| 6 | | | $C_3H_8/O_2/N_2$ | | |

# 3. Results and discussions

## 3.1 Compressibility factors

From Eqs.7 – 9, it is obvious that the *P-T-V* relationship is an important factor influencing real-fluid properties in a reacting process. This relationship can be described by compressibility factors, as shown by Eq. 1, which describe the deviation of real fluid behavior from the ideal fluid behavior (i.e., the real-fluid effect).

Based on the high-order mixture Virial EoS derived in Section 2.1, compressibility factors of three mixtures listed in Table 2 are computed at wide conditions. The results are summarized in Fig. 3. Immediately seen from Fig. 3 are the strong real-fluid effects observed for all the mixtures, which increases with increasing pressure and decreasing temperatures, highlighting the need to investigate real-fluid effects on supercritical combustion at low temperatures. The compressibility factors computed using Virial EoS reach unity (i.e., ideal gas) at 1 bar, irrespective of temperatures, indicating the sufficiency of Virial EoS over sub-, trans- and super-critical conditions. The $CH_3OH/O_2/N_2$ mixture exhibits the highest compressibility factors, with the highest value of 1.75 obtained at T=500 K and P=1000 bar, due possibly to the stronger polarity of $CH_3OH$ molecules [34]. As $N_2$ comprises the major part of the mixture (i.e., 90%) and the polarity is weaker in the other two mixtures, the compressibility factors are nearly the same for $CH_3OCH_3/O_2/N_2$ and $C_3H_8/O_2/N_2$ mixtures.

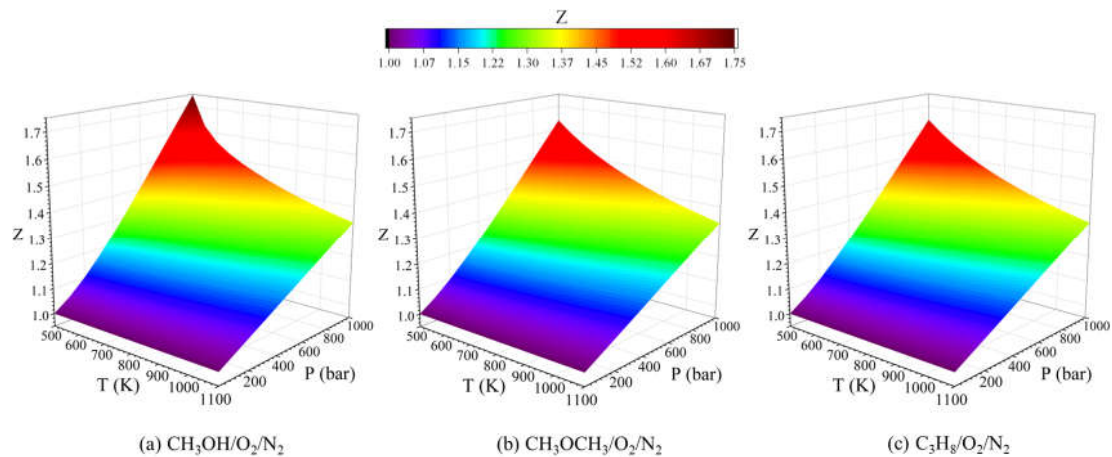

**Figure 3. Compressibility factors of three mixtures (Table 2) at a wider range of temperature and pressure calculated by high-order mixture Virial EoS.**

## 3.2 Supercritical JSR modeling versus experimental data

Supercritical JSR modeling is first conducted for $CH_3OH$, $CH_3OCH_3$, and $C_3H_8$ oxidations at the conditions specified in Table 1. The simulated species profiles are summarized in Fig. 4 – 6 for the fuel and key intermediate species, along with the experimental measurements collected from [31-33], whenever applicable.

For the $CH_3OH/O_2/N_2$ mixtures, as shown in Fig. 4, both ideal and high-order Virial EoS captures the trends of mole fraction change with temperature, although they underestimate the extent of oxidation. This is consistent with all the equivalence ratios studied. Compared with the ideal EoS, the mole fractions calculated by high-order Virial EoS are nearly identical, except for CO, $CO_2$, $CH_2O$ and $H_2$. Specifically, at 100 atm, the real fluid effects lead to lower production of CO, $CH_2O$ and $H_2$, while higher production of $CO_2$. This is most obvious at the high-temperature end, e.g., 850 K.

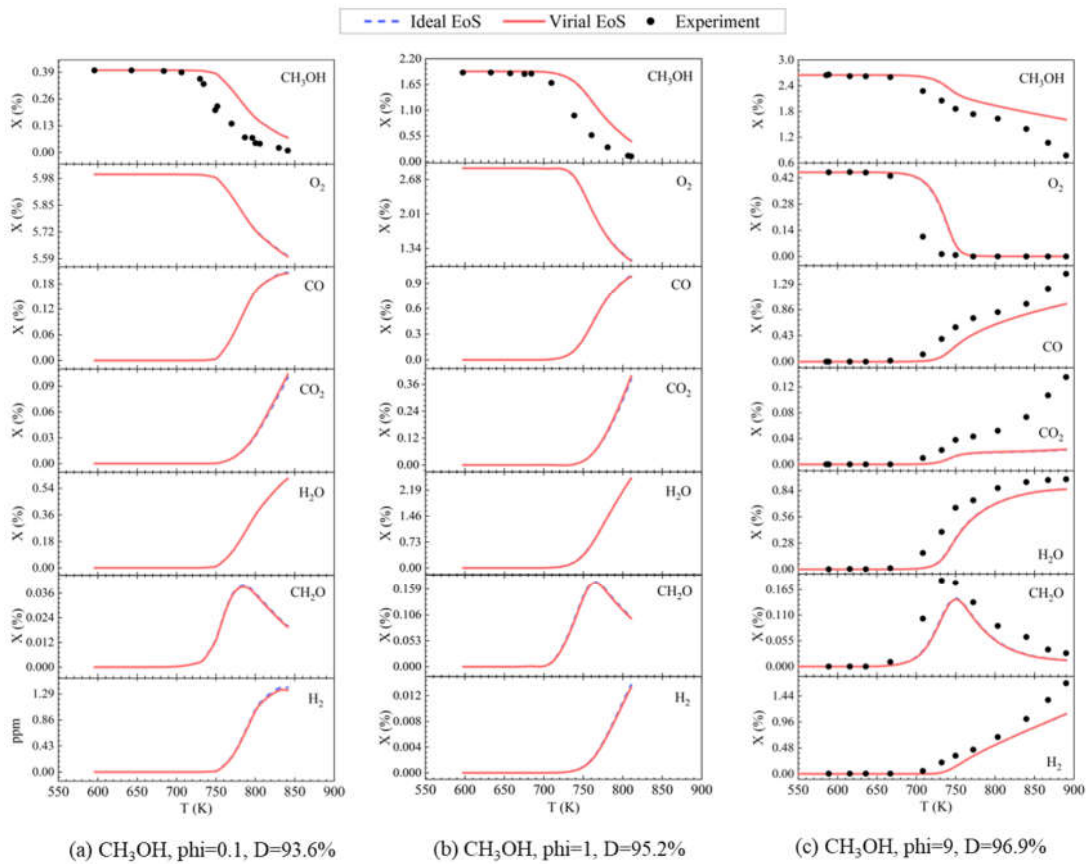

**Figure 4. Simulated species profiles during $CH_3OH/O_2/N_2$ oxidation from supercritical JSR modeling based on high-order Virial EoS and ideal EoS at P=100 atm and T=534-889 K, with comparisons against experiment data [31].**

For the $CH_3OCH_3/O_2/N_2$ mixtures, as shown in Fig. 5, the qualitative trends of real fluid effects on simulated species profiles are very similar to those observed for the $CH_3OH/O_2/N_2$ mixtures (c.f. Fig. 4), with lower production of CO and $CH_2O$ and higher production of $CO_2$. Nevertheless, the quantitative differences between the simulation results based on the high-order Virial EoS and the ideal EoS become more pronounced than those observed in Fig. 4. This is most obvious with $CH_2O$, where the mole fractions of $CH_2O$ at temperatures above 750 K are decreased by real fluid effects, leading to somewhat better agreements with the experimental measurements. Also, the influences from real fluid effects seem to be increased toward higher equivalence ratios.

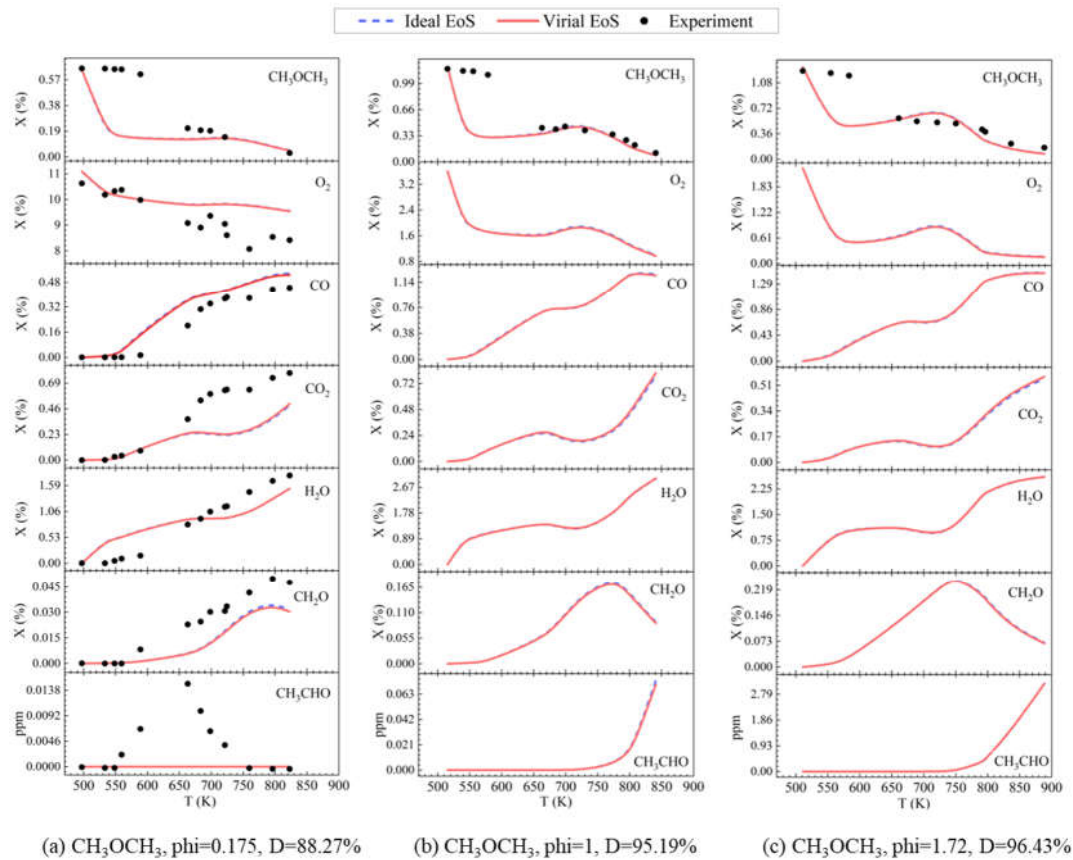

(a) CH$_3$OCH$_3$, phi=0.175, D=88.27%  (b) CH$_3$OCH$_3$, phi=1, D=95.19%  (c) CH$_3$OCH$_3$, phi=1.72, D=96.43%

**Figure 5. Simulated species profiles during CH$_3$OCH$_3$/O$_2$/N$_2$ oxidation from supercritical JSR modeling based on high-order Virial EoS and ideal EoS at P=100 atm and T=497-888 K, with comparisons against experiment data [32].**

For the C$_3$H$_8$/O$_2$/N$_2$ and C$_3$H$_8$/O$_2$/N$_2$/CO$_2$ mixtures, as illustrated in Fig. 6, the impacts of real-fluid effects on the simulation results become more obvious. Specifically, there is a clear and greater shift in fuel species profile (i.e., C$_3$H$_8$) in the temperature range of 650 – 700 K, where greater consumption rates of C$_3$H$_8$ display in the NTC (negative temperature coefficient) region under the influences of real-fluid effects. There is also a clear shift in the CH$_2$O and CH$_3$CHO profiles at temperatures of 650 – 750 K (see Fig. 6a and 6c), which is less significant for the CH$_3$OH/O$_2$/N$_2$ mixtures (i.e., Fig. 4) and the CH$_3$OCH$_3$/O$_2$/N$_2$ mixtures (i.e., Fig. 5). It is also interesting to see that, with incorporating real-fluid effects, the production of CH$_3$CHO and CH$_2$O is promoted for the C$_3$H$_8$/O$_2$/N$_2$ and C$_3$H$_8$/O$_2$/N$_2$/CO$_2$ mixtures (e.g., Fig. 6a), whereas these are inhibited for the CH$_3$OH/O$_2$/N$_2$ mixtures (e.g., Fig. 4a) and the CH$_3$OCH$_3$/O$_2$/N$_2$ mixtures (e.g., Fig. 5a). This highlights the diversified impacts of real-fluid effects on oxidation characteristics, which highly depend on the type of mixture studied. Adequately addressing this diversity requires adequately representing mixture-specific intermolecular interactions with explicit molecular identity, which is challenging to achieve with empirical methods, while can be sufficiently characterized with the framework proposed herein.

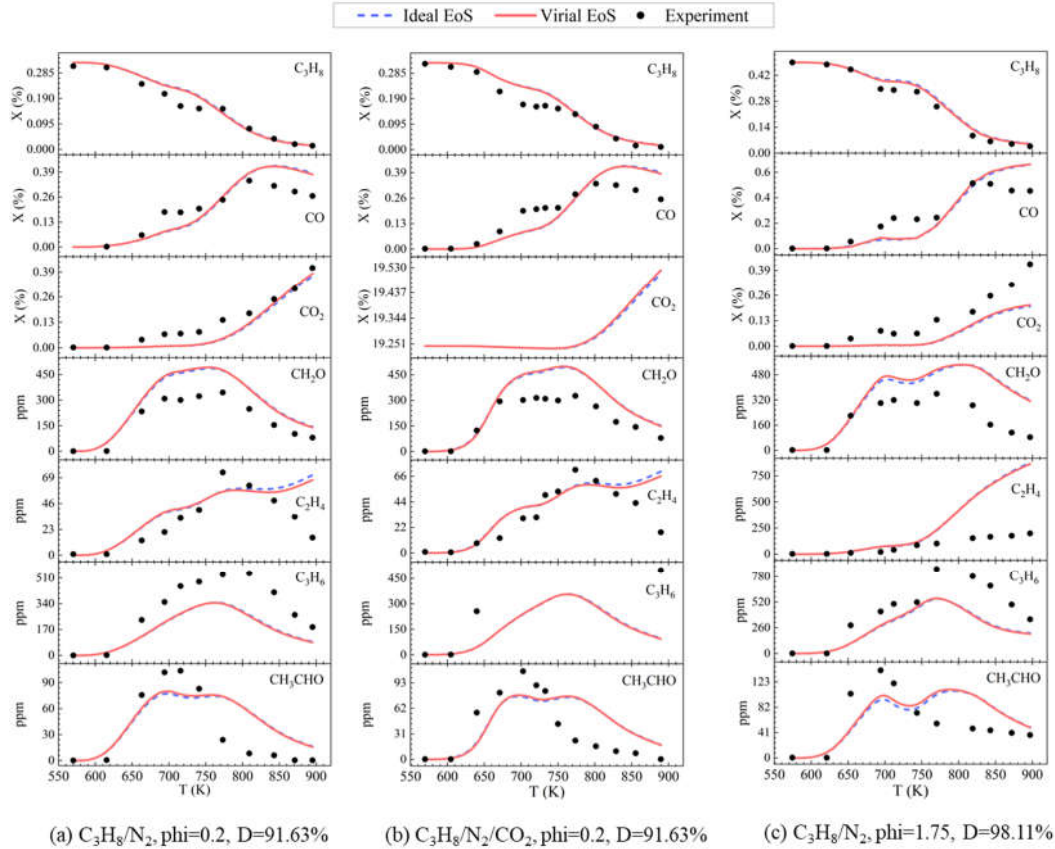

**Figure 6.** Simulated species profiles during $C_3H_8/O_2/N_2$ and $C_3H_8/O_2/N_2/CO_2$ oxidation from supercritical JSR modeling based on high-order Virial EoS and ideal EoS at P=100 atm and T=569-895 K, with comparisons against experiment data [33].

Although Figs. 4-6 witness a gap between the simulated species profiles based on the ideal and high-order Virial EoS, the magnitude of the difference is still limited regardless of temperature, equivalence ratios and dilutions. This can be explained by the compressibility factors presented in Fig. 3. As can be seen from Fig. 3, the compressibility factors for all the mixtures at 100 atm are only slightly above 1.0, indicating the insignificant real-fluid behaviors at this pressure condition. Nevertheless, these results for different mixtures clearly demonstrate that the high-order Virial EoS developed based on *ab initio* intermolecular potentials shows superiority over the ideal EoS in replicating the real-fluid effects when predicting oxidation characteristics in supercritical JSR. These trends are caused by molecular behavior in essence and are discussed in detail in the following section.

## 3.3 Real-fluid effects over wider temperatures and pressures

To better show the real-fluid effects on supercritical oxidation in JSR, further simulations are conducted for the mixtures listed in Table 2 at the pressure of 1000 bar. Immediately seen in Fig. 7 is the considerable impact of real-fluid effects on the oxidation processes. For all mixtures, fuel consumption is promoted by real-fluid effects at lower temperatures (e.g., <650 K for $C_3H_8$ in Fig. 7c), while at higher temperatures, the fuel consumption rate becomes smaller. Generally speaking, with real-fluid effects considered, mole fractions of the final products (e.g., $CO_2$ and $H_2O$) become higher, whereas those of the key intermediates (e.g., $CH_3CHO$, $CH_2O$, CO and $H_2$) become lower. Compared with the results based on ideal EoS, the species profiles for these key intermediates change both quantitatively and qualitatively, indicating that the oxidation pathways and their fluxes are significantly affected by real-fluid behaviors. For instance, due to the advanced onset of $C_3H_8$ oxidation, as can be seen from Fig. 7c, the formation of the primary derivatives such as $C_3H_6$ and secondary derivatives such as $CH_2O$ starts earlier toward lower temperatures, as compared to the results without considering real-fluid effects.

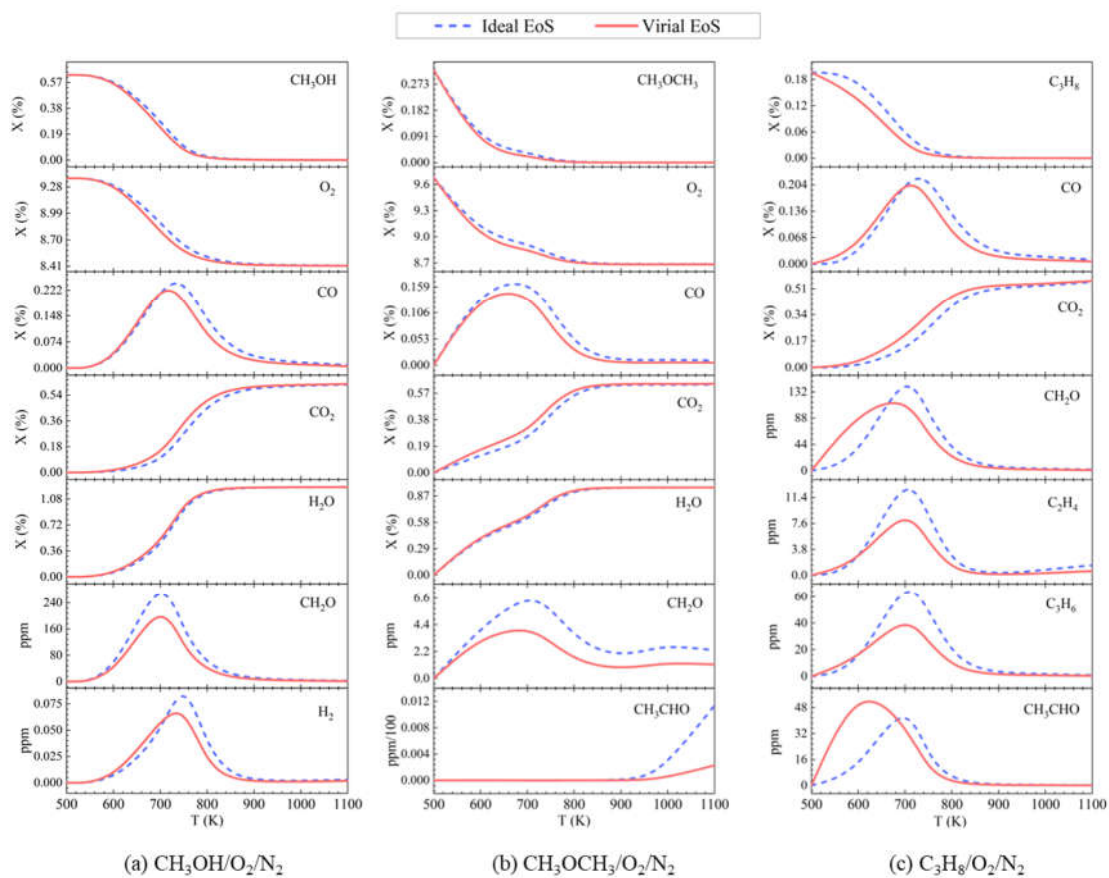

**Figure 7.** Simulated species profiles during the oxidation of three mixtures from supercritical JSR modeling based on high-order Virial EoS and ideal EoS at P=1000 bar and T=500-1100 K.

The large and non-monotonic changes in supercritical oxidation species profiles necessitate the investigation of real-fluid effects in JSR over wider ranges of temperature and pressure conditions. Therefore, supercritical JSR simulations are further conducted at temperatures from 500 to 1100 K and pressures from 100 to 1000 bar. The normalized change in species mole fraction (either in percent or ppm) in comparison to the simulation results based on ideal EoS is determined, which is computed as $(X_{IG} - X_{VR})/X_{IG}$, where $X_{IG}$ and $X_{VR}$ are the simulated mole fractions using the ideal and high-order Virial EoS, respectively. A positive (negative) value of the normalized change indicates suppressed (promoted) production or promoted (suppressed) consumption of the corresponding species. The results are summarized in Fig. 8, where the normalized change is presented in a 2-D temperature-pressure space. The contours are truncated when the absolute mole fraction at a condition is less than 1% of the maximum mole fraction for the specific species, as the small absolute values for $X_{IG}$ tend to give unphysical values for the normalized change.

The results in Fig. 8 obviously confirm that the impacts of real-fluid effects are significant while diversified, with a strong dependence on temperature (non-linear), pressure (somewhat linear) and mixture. Stronger influences are observed at higher pressures, low to intermediate temperatures, and for mixtures containing heavier fuels (e.g., $C_3H_8$), as can be interpreted from the expanded red regions in Fig. 8. In general, real-fluid effects promote fuel reactivity, leading to earlier onset of fuel oxidation and higher fuel consumption rates at lower temperatures. This leads to the promoted production of primary fuel derivatives. These trends agree with those reported in [20] where the real-fluid effects were also found to promote autoignition reactivity. As fuel oxidation is promoted, the production of final combustion products is promoted in general, as can be seen from the blue regions (i.e., negative changes) in the subplots for $CO_2$ and $H_2O$. This implies a prolonged oxidation process under the influences of real-fluid effects, which is confirmed in Fig. 7c, where the production and consumption of the key intermediates (e.g., $CH_3CHO$) span over wider temperature ranges than those of the ideal case. Over the conditions studied, real-fluid effects can lead to >50% faster fuel consumption and >20% higher production of products.

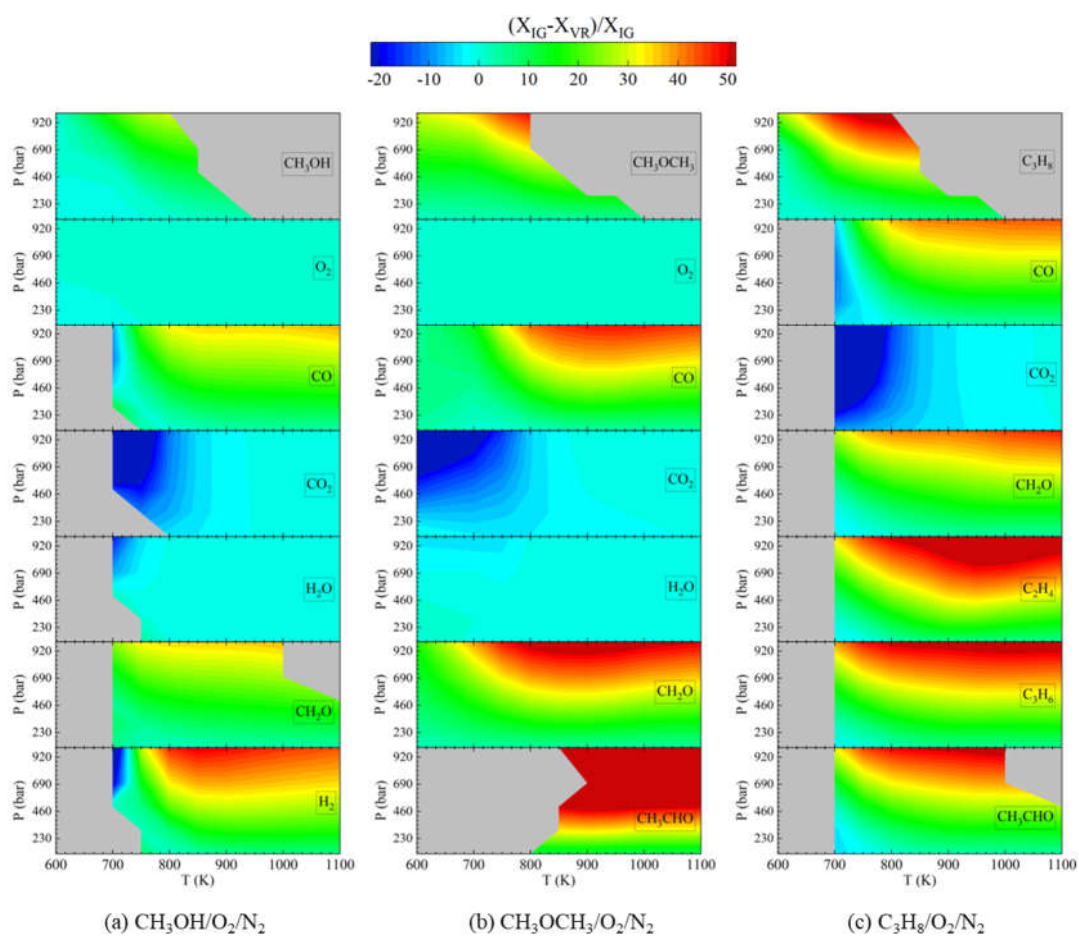

**Figure 8. Deviation of simulated mole fractions calculated based on high-order Virial EoS from those calculated based on ideal EoS in a supercritical JSR at P=100-1000 bar and T=600-1100 K.**

# 4. Conclusions and recommendations

In previous studies concerning supercritical oxidation in JSR, modeling of these experiments has been mainly conducted assuming ideal gas behaviors, without considering the real-fluid effects on the oxidation characteristics. Such assumptions have now been proved in this study to be inappropriate for interpreting high-pressure JSR experiments. This is achieved through a newly proposed framework for supercritical JSR modeling, which couples *Ab initio* intermolecular interactions, high-order Virial EoS and real-fluid oxidation conservation laws for the first time. The proposed framework enables a unique lens into the real-fluid effects on supercritical oxidation processes, which is demonstrated with significantly improved adequacy in simulating supercritical oxidation processes in JSRs.

Through comprehensive simulation campaigns, real-fluid effects are found to greatly affect oxidation characteristics in JSR, with strong promoting effects exhibiting on fuel oxidation reactivity. These real-fluid effects display strong non-linear, somewhat linear and complex dependence on temperature, pressure and mixture compositions, respectively, with stronger influences observed at higher pressures, low to intermediate temperatures, and for mixtures containing heavier fuels.

The significant influences of real-fluid behaviors on JSR oxidation characteristics emphasize the need to adequately incorporate these effects for future modeling studies in JSR at high pressures. Without doing so, significant uncertainties or even unrealistic modeling results will be introduced, leading to misunderstanding or faulty interpretation of modeling results. For instance, the species profiles can change quantitatively by approximately 65% in absolute mole fraction (e.g., $C_3H_6$ in Fig. 7c) and qualitatively with completely different production or consumption trends (e.g., $CH_3CHO$ and $CH_2O$ in Fig. 7c). Under such conditions, validating chemical kinetic mechanisms against supercritical JSR experiments without adequately accounting for the real-fluid effects can make the validation work meaningless, given that the typical levels of measurement uncertainty in JSR are on the order of ±20%. Fortunately, this can now be achieved through the framework proposed in this study.

# Acknowledgments

This material is based on work supported by the Research Grants Council of Hong Kong Special Administrative Region, China, under PolyU P0046985 for ECS project funded in 2023/24 Exercise and P0050998, and by the Natural Science Foundation of Guangdong Province under 2023A1515010976.

# Declaration of Competing Interests

The authors declare no competing interests.

# Additional information

Details of the generating the high-order mixture Virial equation of state with related validations and comparisons are provided in the Supplementary Information. The data shown in Figs. 3-8 are available upon request from the corresponding author. These results can be reproduced using the governing equations derived in the Supplementary Information and publicly available codes and mechanisms, such as CANTERA and NUIGMech1.1 mechanism.